\documentclass[aps,prl,twocolumn,showpacs,superscriptaddress,groupedaddress]{revtex4-1} 
\usepackage{graphicx}  % needed for figures
\usepackage{dcolumn}   % needed for some tables
\usepackage{bm}        % for math
\usepackage{amssymb}   % for math
\usepackage{xcolor}
\usepackage{graphics} 
\usepackage{hyperref}

\usepackage{epsfig} 
\input{epsf}

\def\beq{\begin{equation}}
\def\eeq{\end{equation}}
\def\be{\begin{equation}}
\def\ee{\end{equation}}
\def\bea{\begin{eqnarray}}
\def\eea{\end{eqnarray}}

\begin{document}

\title{Repulsive vector interaction as a  trigger for the  non-conformal peak in $V_s^2$}
 
\author{Marcus BENGHI PINTO} \email{marcus.benghi@ufsc.br}
\affiliation{Departamento de F\'{\i}sica, Universidade Federal de Santa
  Catarina, Florian\'{o}polis, SC 88040-900, Brazil}

 \begin{abstract} 
Considering  the NJL  model with a repulsive vector channel, parametrized by $G_V$, I show that one may generate a non-monotonic behavior  for the speed of sound which  peaks at $V_s^2 > 1/3$.  This can be achieved by assuming  $G_V$ to be density dependent so that the resulting  EoS  is stiff/repulsive at low densities  and  soft/non-repulsive at high densities. The interpolation between the two regimes happens through a cross-over which takes place after the first order chiral transition sets in. The model  explicitly   shows that a non-conformal peak in $V_s^2$ is not in tension with the QCD trace anomaly being positive at all densities, supporting recent claims in this direction. A brief discussion on how the running coupling may affect the mass-radius relation is carried out in the context of simple non-strange quark stars. 

\end{abstract}

\maketitle
\section{Introduction}
 Understanding how strongly interacting matter behaves at high densities and low temperatures is crucial to describe neutron  stars (NS)  which represent the only laboratory where cold and dense quantum chromodynamics (QCD) can be currently tested. This can be understood by recalling that at the present time most  relativistic heavy ion collisions  experiments are devoted to study  hot and moderately compressed hadronic matter. At the same time, this regime is not yet fully accessible to lattice simulations. On the theoretical side, the recent discovery  of NS whose estimated masses are about twice the value of the solar mass \cite{astro1, astro2,astro3} implies that the existing equations of state (EoS) need to be further improved in order to present higher stiffness. An observable which can describe the stiffness of matter is the  speed of sound, $V_s$. At vanishing temperatures this quantity can be evaluated from the knowledge of the baryon number density, $n_B$.  Thermodynamic stability and causality allow the value of the speed of sound to be within a generous range,  $0 \le V_s^2 \le 1$. Fortunately,  this   large  uncertainty  can be  further reduced by considering the extreme limits of very low and very high  densities. In the  first case, $n_B \lesssim n_0$ ($n_0=0.16\,{\rm fm}^{-3}$), the EoS can be appropriately described by effective field theory models \cite {chpt1,chpt2,chpt3,chpt4} which predict small values,  $V_s^2 << 1$. At the other extreme, $n_B \gtrsim 40\, n_0$, where the   EoS can be described by  perturbative QCD (pQCD)  \cite{pqcd1,pqcd2}  the  speed of sound converges towards the  conformal value,  $V_s^2 = 1/3$.  Between these two limits, the EoS cannot be derived from  {\it ab initio} evaluations so that  the value of the speed of sound within compressed baryonic matter remains essentially unknown (Ref. \cite {kojo} offers a detailed discussion on the possible scenarios). While some authors \cite {paulo,silva} advocate the existence of an universal bond, $V_s^2 < 1/3$, the measurements performed in Refs. \cite{astro1, astro2,astro3} and the theoretical predictions on the maximum (gravitational) mass performed in Refs.  \cite{measure1, measure2, measure3,measure4,measure5} favor a stiff EoS with $V_s^2 \gtrsim 1/3$  at $n_B \gtrsim n_0$. In this case, recent simulations \cite {sinansimulation} indicate that the most probable scenario is the one in which $V_s^2$ is a non-monotonic function of $n_B$, which in turn suggests  the existence of at least one local maximum where $V_s^2 > 1/3$.  The results presented in Ref. \cite {sinansimulation} were obtained
from a statistical analysis, based on models consistent not only with
nuclear theory and perturbative QCD, but also with astronomical
observations. Moreover, the possible existence of a non-conformal peak in $V_S^2$ is supported by a large number of applications employing frameworks such as quarkyonic matter \cite {sanjay, theo1, theo2}, models for dense QCD \cite {  theo5, theo6,theo7, theo8, theo9, theo11, theo12}, as well as  models based on the gauge/gravity duality \cite {theo13, theo14, theo15}, hadron percolation threshold \cite {michal}, and Bayesian inference methods \cite{weisesimulation} among others. Regarding the present application, which deals with a pure quark EoS, it is important to mention that the local maximum observed in Refs. \cite{theo5,theo6,theo7,theo8}  comes from the hadronic part of a hybrid EoS.

To emphasize the importance of $V_s^2$ it is worth mentioning that changes in its slope can also provide important information related to  the phase transitions and cross-overs that can take place within dense baryonic matter. Therefore, given its essential role  in the description of NS, the  speed of sound recently became the object of intense research. Since first principle evaluations are still not accessible in the relevant density regime some authors have chosen a more pragmatic strategy where the description is carried out through simple models (incorporating some  {\it ansatz})  which supposedly capture the physics necessary to describe the non-conformal peak \cite{sanjay,fukushimatrace}. Inspired by Refs. \cite{sinansimulation,fukushimatrace}, this work aims to provide an alternative framework, based on the  Nambu--Jona-Lasinio model (NJL) \cite{njl}, which may help to answer (even if partially) some of the following questions: i) what is the physical origin of the non-conformal peak?, ii) what type of function is $V_s^2(n_B)$?, iii) is  the trace anomaly always positive in dense matter?, iv) the existence of a non-conformal peak implies the existence of other phase transitions or cross-overs? Here, I suggest that the NJL model with a density dependent repulsive interaction, parametrized by the coupling $G_V$,  provides a solid framework to analyze the non-conformal behavior displayed by $V_s^2$. After proposing an {\it ansatz} to describe how $G_V$ runs with the quark chemical potential, $\mu$, I find that $V_s^2$ indeed has a non-monotonic behavior with a peak at $n_B\simeq  3.25 \, n_0  $. The quark susceptibility shows that the expected (chiral) transition happens at $n_B\simeq 2.5 \,n_0  $ but a novelty shows up at  $n_B \simeq 6\, n_0$ when a cross-over from a  stiff EoS to a  soft one takes place. As a result, the presence of  a non-conformal peak is not in tension with a positive trace anomaly, $\Delta$, (at all densities)  in agreement with the recent conjecture made in Ref. \cite {fukushimatrace}. Finally, a naive application to the case of non-strange stars suggests that  $G_V=0$ and  $G_V(\mu)$ observe the bond $\Delta >0$, contrary to the fixed $G_V$ case (which gives the stiffest EoS).

\section{Model set up}
 In order to account for nuclear repulsion, Nambu suggested \cite{nambuvec} that the Yukawa potential should receive a  vector  contribution. The need to consider this contribution,  when describing nuclear matter by means of a quantum field theory, was later recognized by  Walecka \cite{walecka1}.  On the other hand, the original NJL model considered here was originally proposed in terms of  scalar and  pseudo-scalar channels parametrized by  $G_S$ before Koch et al. \cite {volker}  introduced a repulsive vector channel, parametrized by $G_V$, in order  to account for stability. In this case, the extended $N_f=2$ theory  can be  described by 
\begin{equation}
{\cal L}_{NJL} = {\cal L}_0
 + G_S [ ({\overline \psi} \psi)^2 - ({\overline \psi} {\vec \tau} \gamma_5 \psi)^2] - G_V ({\overline \psi} \gamma^\mu \psi)^2 \;,
\end{equation}
where ${\cal L}_0 = {\overline \psi}(i\gamma_\mu\partial^\mu-m)\psi$.
To assure rotational invariance only the zeroth component of the vector channel contributes so that, at the mean field level, the chemical potential gets shifted as $\mu \to \mu - 2G_V n$, with  $n=3n_B$ representing the quark number density, while the pressure receives a contribution proportional to $G_V n^2$  \cite{buballa, fukushima1,fukushima2}. In $3+1\,d$ the NJL interactions are described by irrelevant operators and   the  couplings turn out to have  canonical dimensions [-2], implying that the model is non-renormalizable. In most cases the divergent integrals are  regularized by a sharp cut-off, $\Lambda$, which is also the procedure adopted here. This new ``parameter" is then fixed, together with $G_S$ and the quark current masses, by requiring  the model to reproduce the phenomenological values of $f_\pi$, $m_\pi$ and $\langle {\overline \psi} \psi \rangle$ at $T=\mu=0$. Here, for simplicity I consider $m_u=m_d\equiv m$ and   then adopt the following parametrization:  $m= 5.6 \, {\rm MeV}$, $\Lambda = 587.9\, {\rm MeV}$ and $G_S \Lambda^2 = 2.44$ \cite {buballa}. 
However, fixing $G_V$ poses and
additional problem since this quantity should be fixed using the $\rho$ meson mass which, in general, happens to be
higher than the maximum energy scale set by $\Lambda$. In this situation, most authors adopt values between $0.25 G_S$ and  $0.5 G_S$ (see Ref. \cite {tulio} for more details). The present work is totally based on the possibility that the value of $G_V$ {\it varies} with $\mu$ just like  $\alpha_s$  in pQCD applications. A crucial difference is that the $\alpha_s$ running is dictated by {\it ab initio} evaluations of the QCD $\beta$ function before $\mu$ gets related to the ${\overline {\rm MS}}$ renormalization scale. In the NJL case one alternative is to use plausible physical arguments  in order to obtain an {\it ansatz} which gives a physically appealing running. With this aim, let us start by imposing that, at low-$\mu$,   $G_V(\mu)$ reproduces the result predicted by  Sugano et al. \cite {sugano}, $G_S/3$ (although this value has been obtained  with the more sophisticated entangled Polyakov-NJL model it remains within the canonical range, $G_V = 0.25-0.5 \,G_S$).  The next step is to determine the intermediate scale at which the chiral transition occurs. By using the adopted parametrization and standard mean field evaluations one finds that the quark effective mass value at zero density is $M(0)=400\;{\rm MeV}$ \cite{buballa}. Since chiral symmetry will be partially restored at $\mu \sim M(0)$ we can further impose that $G_V $ be approximately constant from $\mu=0$ to $\mu=M(0)$ so that the usual results for the first order chiral transition obtained with a fixed coupling are preserved. We next require $G_V(\mu) \to 0$ at a larger scale, such as $\mu \simeq \Lambda$. Expecting the decrease in $G_V(\mu)$ to be more intense between $\mu=M(0)$ and $\mu=\Lambda$ we finally require $G_V(\mu_0) = G_V(0)/2$ at a particular scale, $\mu_0 = [M(0)+\Lambda]/2$. Then, it is not difficult to foresee that the required form of $G_V(\mu)$ is reminiscent of the Woods-Saxon potential. Namely,
\begin{equation}
G_V(\mu) = \frac{G_V(0)} { 1 + e^{(\mu - \mu_0)/\delta}} \;,
\label{run}
\end{equation}
where $\mu_0=  500 \, {\rm MeV}$ and $G_V(0)=G_S/3$ .  The ``thickness"  $\delta = 10\, {\rm MeV}$ assures that the drop starting at $\mu=M(0)$ terminates at $\mu = \Lambda$.  It is  obvious from Eq. (\ref{run}) that such running coupling interpolates between the two extrema, $G_V=0$ and $G_V =G_S/3$, which respectively give a softer  and a stiffer  EoS \cite{buballa, fukushima1,fukushima2}. Therefore,   $G(\mu)$  has the potential  to reproduce the expected  non-conformal maximum in $V_s^2$. Note that the {\it ansatz} tacitly implies that after chiral symmetry gets (partially) restored the repulsion among the (bare) quarks becomes negligible  as the density increases. Also, remark that $\delta$ was chosen   so as to give a smooth transition within a narrow $10\,{\rm MeV}$ width since taking $\delta \to 0$ could lead to discontinuities in $V_s^2$ which do not seem to be observed in  Refs.  \cite {sinansimulation,michal,weisesimulation}. With this conservative choice  one can anticipate that the  transition from the repulsive/stiff phase to the non-repulsive/soft phase will be driven by a cross-over. The approach adopted here is similar to the one originally employed by Kunihiro \cite {kunihiro}, who considered  a temperature dependent $G_V$   in order to evaluate quark susceptibilities at high-$T$ (see also Ref. \cite {lorenzo}). 

\section{ Evaluations and numerical results} 

Let us consider the quark number density as representing the fundamental quantity of interest. Then,  at $T=0$, a standard MFA evaluation yields the following  {\it per flavor} result \cite{tulio}
\begin{equation}
n_f = \frac{N_c}{3\pi^2} p_{F,f}^3 \;,
\end{equation}
where the Fermi momentum is $p_{F,f} = \sqrt { {\tilde \mu}_f^2 - M_f^2}$ with ${\tilde \mu}_f = \mu_f  - 2 G_V \sum n_f$. The quark effective mass is given by  $M_f = m - 2 G_S \sum \sigma_f$ where
\begin{eqnarray}
\sigma_f &=& - \frac{N_c}{2\pi^2} M_f \left [ \Lambda p_{\Lambda,f} - M_f^2 \ln \left ( \frac{\Lambda + p_{\Lambda,f}}{M_f} \right)  \right ] \nonumber \\
&+& \frac{N_c}{2\pi^2} M_f  \left [ {\tilde \mu}_f p_{F,f} - M_f^2 \ln \left ( \frac{{\tilde \mu}_f + p_{F,f}}{M_f} \right)  \right ] \,,
\end{eqnarray}
where $p_{\Lambda,f} = \sqrt { \Lambda^2 + M_f^2}$. Having the quark density, $n=\sum_f n_f$, one  can obtain the squared speed of sound from   $V_s^2 = n_B/[\mu_B (d n_B/d\mu_B)]$, where $\mu_B=3\mu$.

\begin{figure}
\centerline{ \epsfig{file=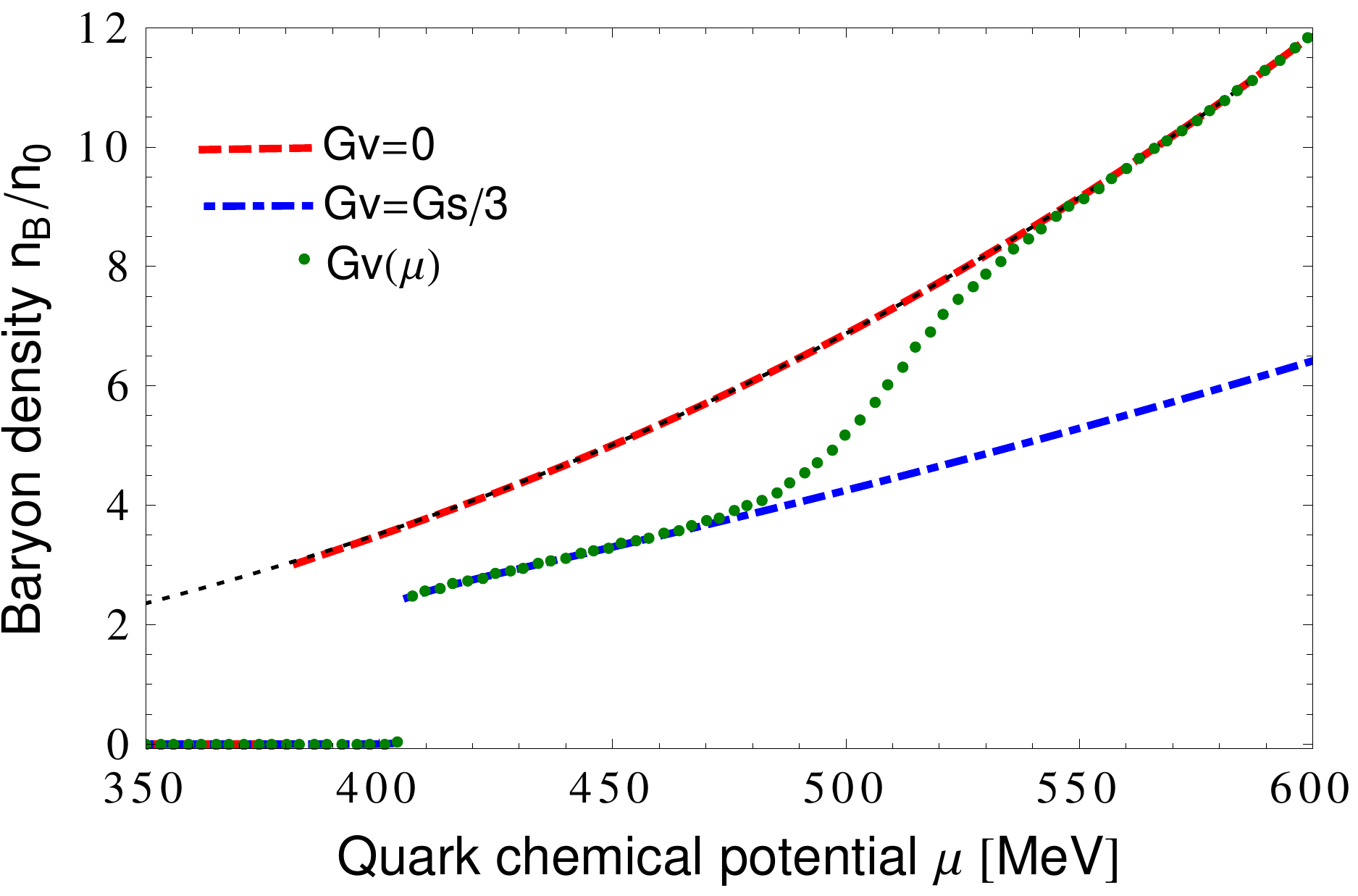,width=0.85\linewidth,angle=0}}
\caption{Baryon density for symmetric quark matter, in units of $n_0$, as a function of the quark chemical potential.  The thin dotted line shows the result for the quark number density $n^{\rm free} = N_c N_f \mu^3/(3\pi^2)$ which corresponds to a gas of free quarks.  }
\label{Fig1}
\end{figure}

\begin{figure}
\centerline{ \epsfig{file=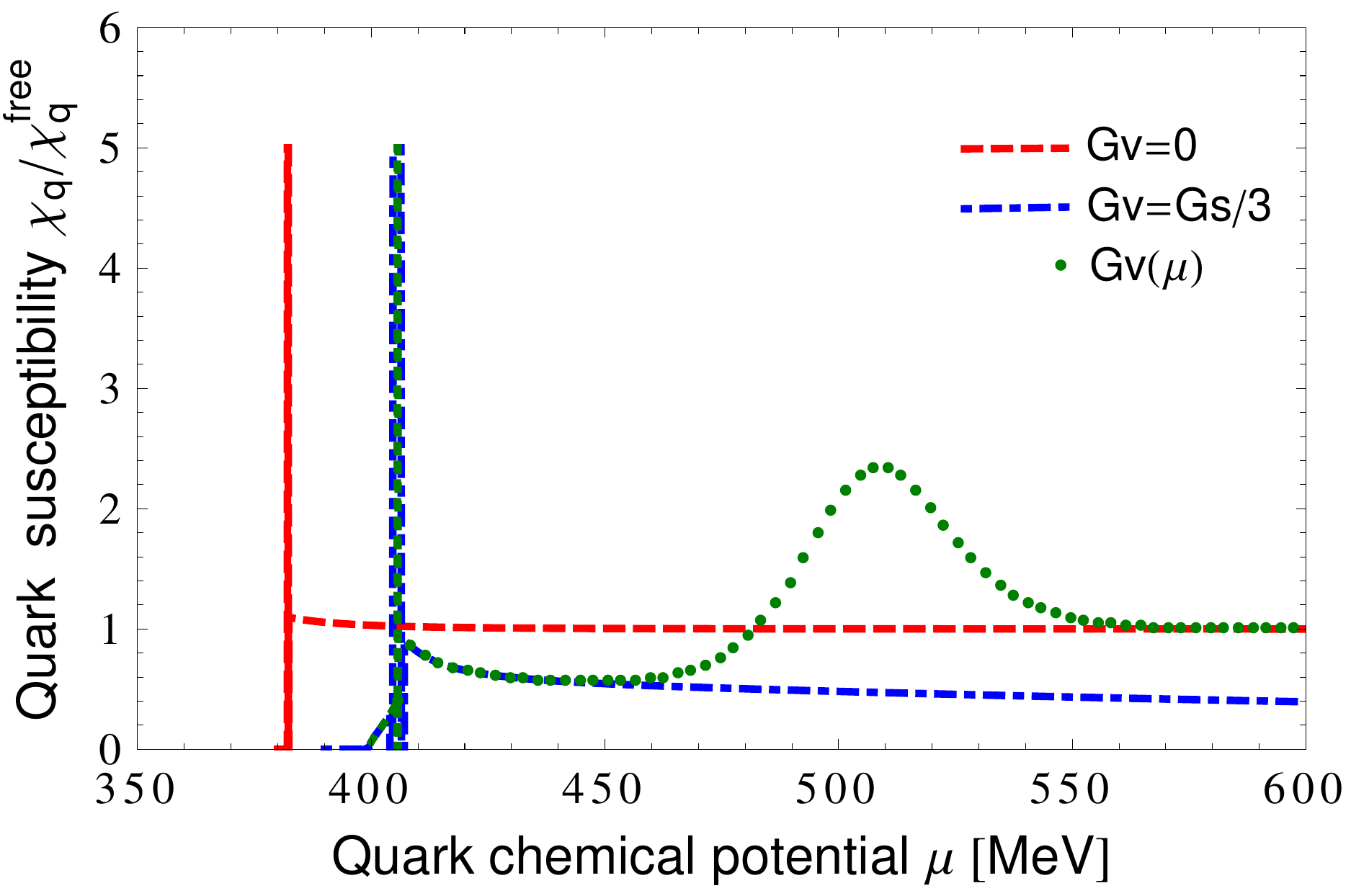,width=0.85\linewidth,angle=0}}
\caption{Quark number susceptibility for symmetric quark matter, normalized by $\chi_q^{\rm free} = N_c N_f \mu^2/\pi^2$, as a function of the quark chemical potential.  After the first order (chiral)  transition  a cross-over can be observed at $\mu= 508.7 \,{\rm MeV}$ for the case $G_V(\mu)$. }
\label{Fig2}
\end{figure}

\begin{figure}
\centerline{ \epsfig{file=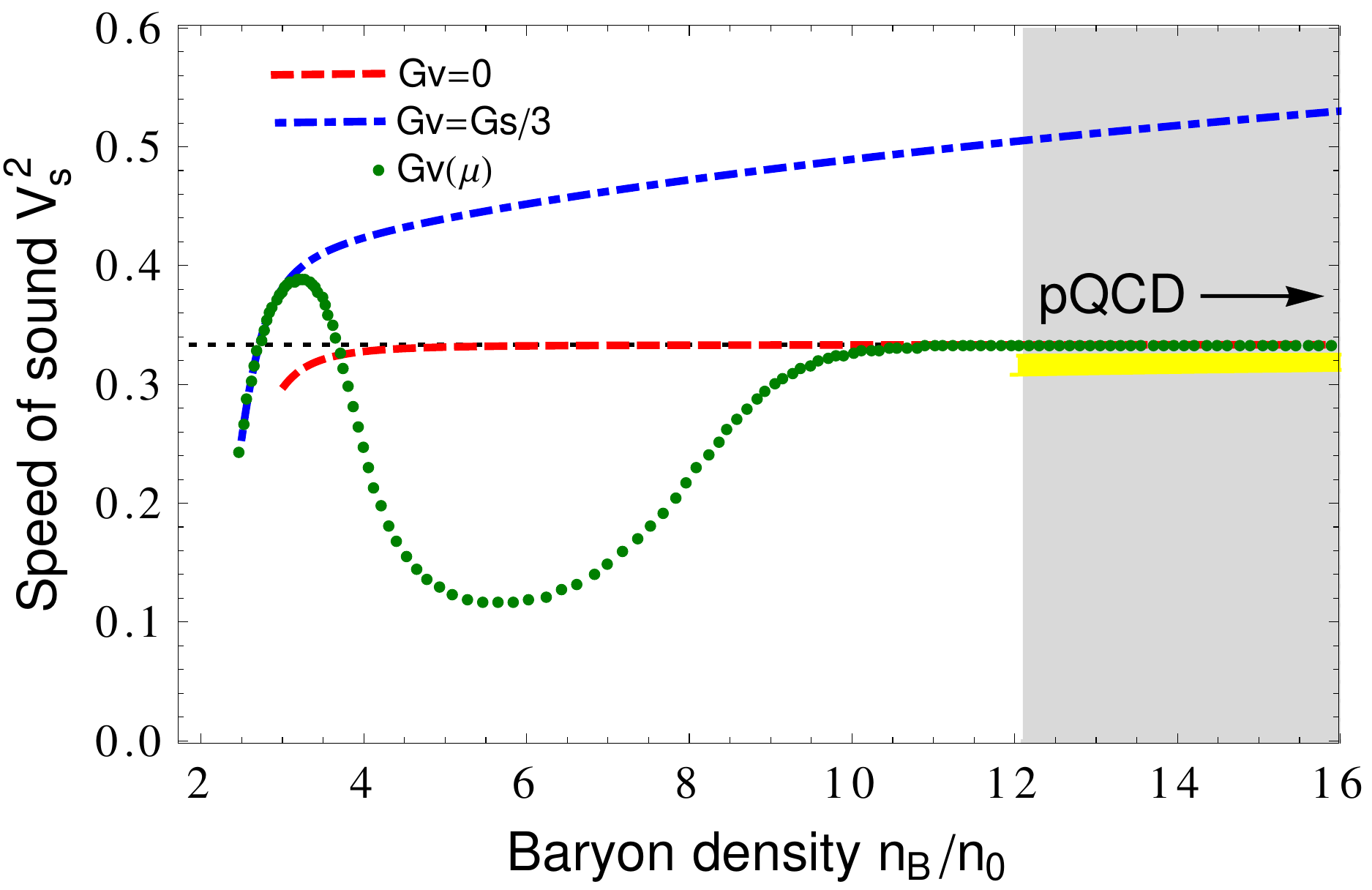,width=0.85\linewidth,angle=0}}
\caption{Speed of sound (squared),  for symmetric quark matter, as a function of $n_B/n_0$. The gray band represents the  $\mu > \Lambda$ region. The light  band corresponds to the pQCD results for $\overline {\rm MS}$ renormalization scales from the central scale, $2 \mu$ (bottom edge), to $4\mu$ (top edge). The thin dotted line represents the conformal result, $V_s^2=1/3$. The peak, $V_s^2=0.39$, occurs at $n_B = 3.25 \, n_0= 0.52\, {\rm fm}^{-3}$.}
\label{Fig3}
\end{figure}

At finite chemical potential and zero temperature, the pressure versus chemical potential relation for quark matter can be obtained from \cite{fukushima2, zong}
\begin{equation}
P(\mu) = P(0) + \int_0^\mu n(\nu) d\nu \,,
\end{equation}
 where $P(0) = $ is the vacuum pressure.  From  $P(\mu)$ one can determine  the energy density, $\epsilon = - P + \mu_B n_B$, the trace anomaly, $\Delta = \epsilon - 3P$, as well as the conformal measure, ${\cal C} = \Delta/\epsilon$. For simplicity let us start by considering the case of symmetric quark matter, $\mu_u=\mu_d\equiv \mu$. Fig. \ref{Fig1} illustrates the baryon density as a function of $\mu$ for $G_V=0$, $G_V= G_S/3$ and $G_V(\mu)$. The figure clearly shows how  $G_V(\mu)$ interpolates between the other two cases predicting that, after the chiral transition, $n_B$  converges to the free gas result. The possible phase transition patterns can be better analyzed by evaluating the quark number susceptibility, $\chi_q = d n /d\mu$. The results displayed in Fig. \ref{Fig2} show that all the three possibilities reproduce the usual first order (chiral) transition which, as expected, is delayed and softened when $G_V\ne 0$ \cite{buballa,fukushima1,fukushima2,tulio}. On top of that, at $\mu=508.73\, {\rm MeV}$ the running coupling induces a cross-over towards the free gas result.  Fig. \ref {Fig3} shows the squared speed of sound as a function of $n_B$. The results obtained with the running coupling indicate that $V_s^2$ exceeds the conformal limit at $n_B \simeq 2.8 n_0$ for both cases in which repulsion is present. However, when $G_V$ is fixed, $V_s^2$ continues to rise monotonically whereas a non-monotonic behavior is displayed by the running coupling which produces a peak, $V_S^2 = 0.39$, at  $n_B =  3.25 n_0 = 0.52 \, {\rm fm}^{-3}$. After that, $V_s^2$ returns to the sub-conformal region and reaches a minimum induced by a cross-over (at $n_B \simeq 6 n_0$) before converging to the conformal  value. The figure also illustrates the pQCD results when the $\overline {\rm MS}$ renormalization scale varies from the ``central" value, $2 \mu$, to $4\mu$. The pQCD predictions were obtained by adapting the $N_f=2+1$ results of Ref. \cite{eduardo} to $N_f=2$. This can be done with ease, owing to the fact that the authors have presented  {\it per flavor} results to the first non-trivial order. Notice that the conjectured coupling running predicts that after peaking at the super-conformal region, $V_s^2$ approaches the conformal value from below, like pQCD, whereas evaluations performed with the {\it hard density loop} resummation \cite{hdl} predict that the approach is  from above. A preliminary  analysis with the {\it renormalization group optimized perturbation theory} resummation \cite {kneur} also indicates that the approach is from below \cite{nos}.  Finally, it should be emphasized that the shape of the  curve generated with  $G_V(\mu)$ resembles some of those recently predicted in Refs. \cite {sinansimulation,michal,weisesimulation}.

\begin{figure}
\centerline{ \epsfig{file=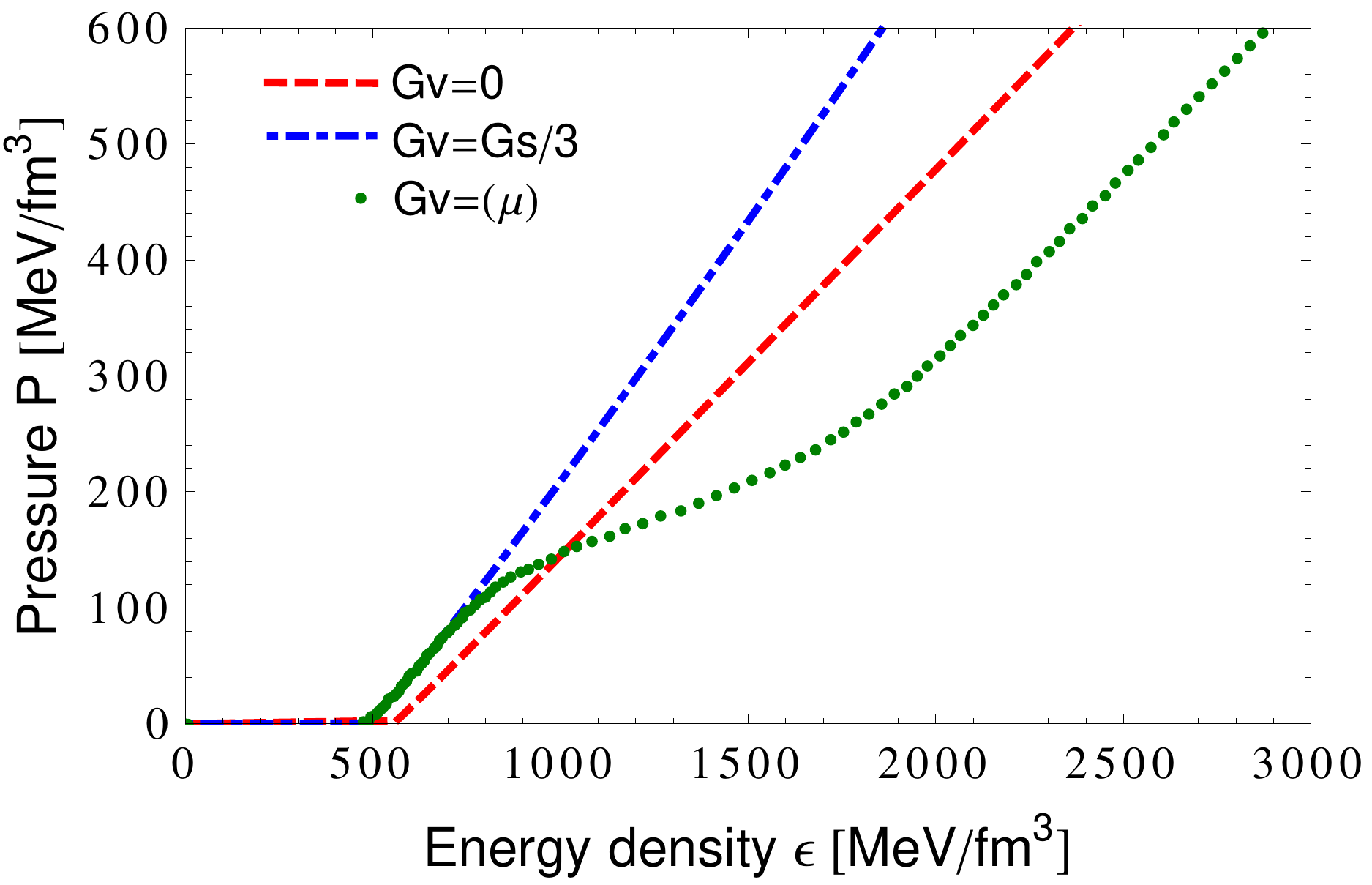,width=0.85\linewidth,angle=0}}
\caption{EoS for  symmetric quark matter. The $G_V(\mu)$ result predicts a stiff EoS at low energies. A cross-over to a softer EoS takes place at $\epsilon = 1286 \, {\rm MeV fm^{-3}}$. The vacuum pressure has been subtracted.}
\label{Fig4}
\end{figure}

Fig. \ref {Fig4} indicates that when  $G_V$ runs with $\mu$ the EoS  is  stiff  for $\epsilon \approx 500-700\, {\rm MeV/fm^{-3}}$. It then becomes very soft  before the cross-over  takes place at $\epsilon  =1286 \, {\rm MeV/fm^{-3}}$ resulting in $dP/d\epsilon \to 1/3$ when $\epsilon  \gtrsim 2000 \, {\rm MeV/fm^{-3}}$ .

\begin{figure}
\centerline{ \epsfig{file=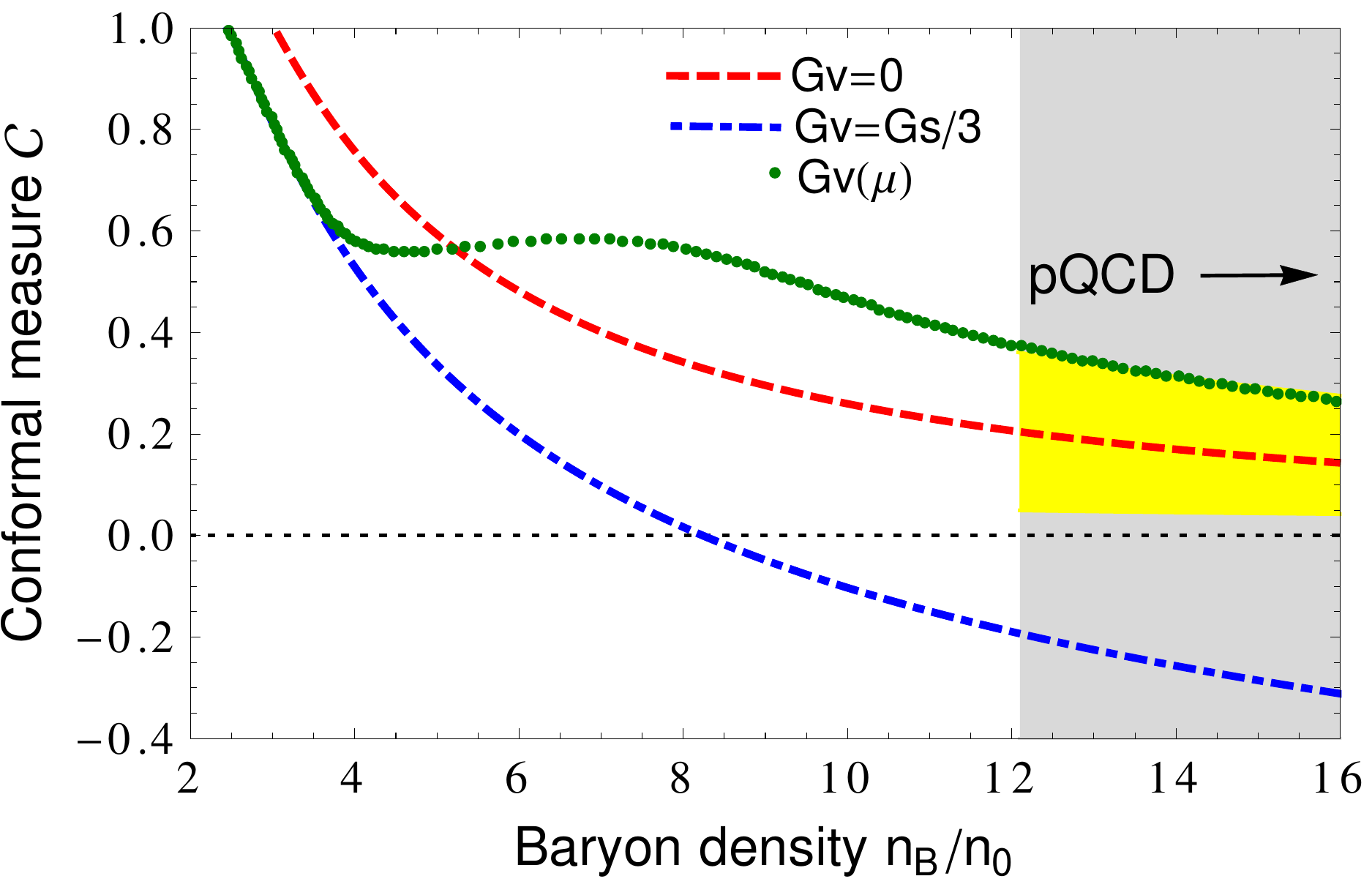,width=0.85\linewidth,angle=0}}
\caption{Conformal measure, for symmetric quark matter, as a function of $n_B/n_0$. The gray band represents the  $\mu > \Lambda$ region. The light  band corresponds to the pQCD results for $\overline {\rm MS}$ renormalization scales from the central scale, $2 \mu$ (top edge), to $4\mu$ (bottom edge). }
\label{Fig5}
\end{figure}

Having in mind a very recent analysis  about  the sign of the trace anomaly \cite{fukushimatrace} let us now investigate how the related conformal measure behaves for the  $G_V$ values considered in this work.  Fig. \ref{Fig5} shows that the fixed $G_V= G_S/3$ produces a maximally stiff EoS which yields a negative $\cal C$ for $n_B \gtrsim 8.3 n_0$. When repulsion is absent, the EoS is softer causing ${\cal C} \to 0$ as $n_B \to \infty$ in conformity with pQCD predictions. At the same time, our running coupling predicts that the cross-over, at $n_B = 6 n_0$, prevents $\cal C$ from diving into the ${\cal C} < 0$ region. It shifts the  high-$n_B$ behavior of the trace anomaly which then converges to zero while remaining positive. Curiously, the NJL with  $G_V(\mu)$ and pQCD give similar results when the latter is evaluated at the $\overline {\rm MS}$ central scale, $2\mu$.

\begin{figure}
\centerline{ \epsfig{file=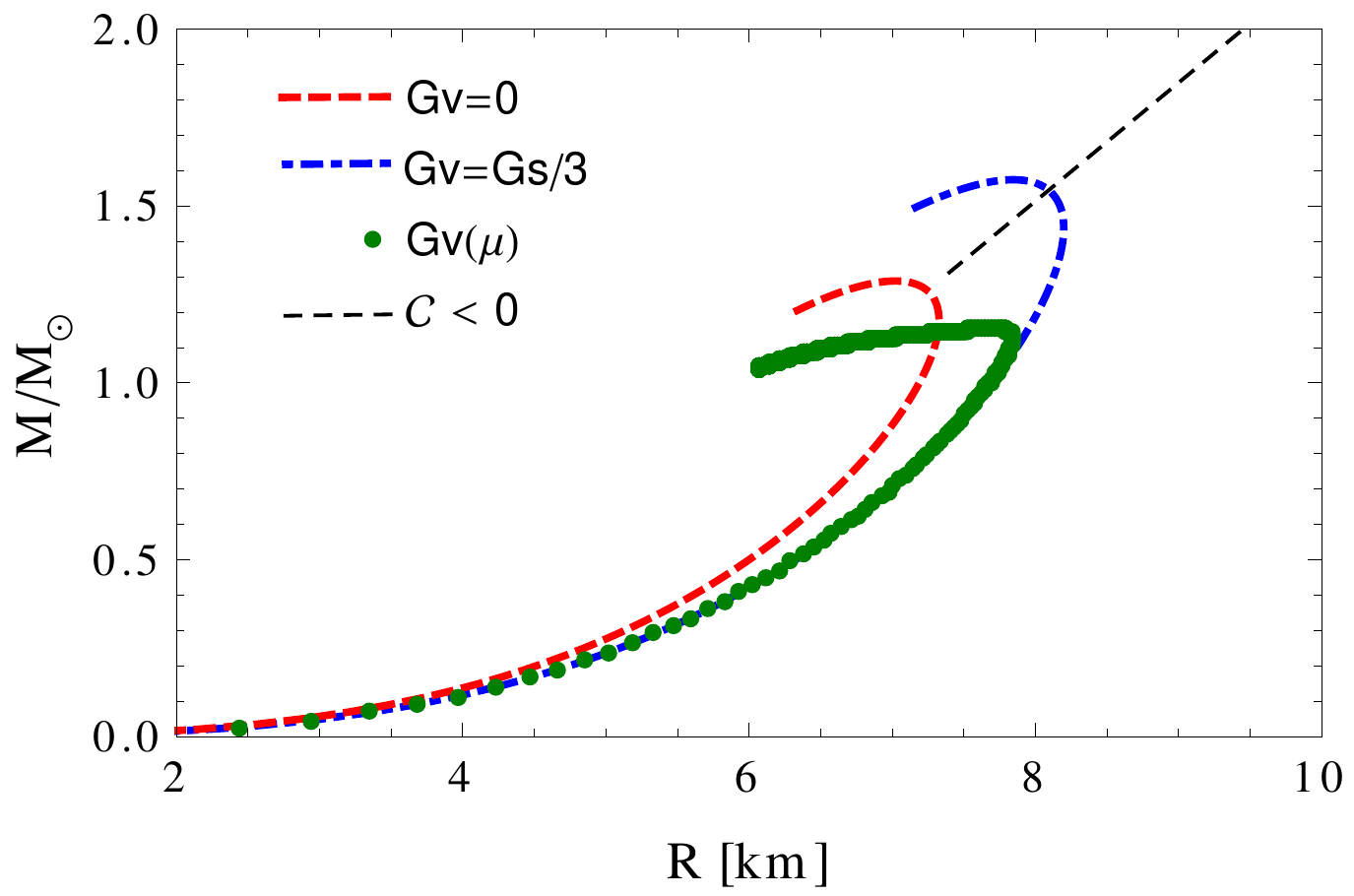,width=0.85\linewidth,angle=0}}
\caption{Mass-radius relation  for a non-strange neutrally charged quark star in $\beta$-equilibrium. The  $M_{\rm max}/R$ ratios are 0.18 for $G_V=0$, 0.19 for $G_V=G_S/3$ and 0.15 for $G_V(\mu)$. The thin dashed line indicates the maximum mass-radius values which could be obtained by using EoS with different fixed $G_V$ values, as in Ref. \cite{jaziel}, which do not observe the ${\cal C }>0$ bond.   }
\label{Fig6}
\end{figure}

Finally, to get a general idea on  how $G(\mu)$  may impact the description of NS let us apply our model to the case of non-strange quark stars. This can be achieved by enforcing $\beta$-stability and charge neutrality upon requiring
$\mu_d - \mu_e = \mu_u \equiv \mu$ and $n_e = (2 n_u - n_d)/3$ with  $n_e = \mu_e^3/(3 \pi^2)$. After solving the TOV equations one obtains the mass-radius relations displayed in Fig. \ref{Fig6} where the results for the standard cases $G_V=0$ and $G_V= G_S/3$ are in agreement with Ref. \cite {jaziel}. The maximum mass-radius ratio obtained with $G(\mu)$ is  0.15 while those obtained with $G_V=0$ and $G_V= G_S/3$ are respectively 0.18 and 0.19. Referring to the discussion carried out in Ref. \cite{fukushimatrace} it is important to mention that the case with fixed $G_V$ does not observe the ${\cal C} > 0$ bond, contrary to the cases $G_V=0$ and $G_V(\mu)$. In the view of modern observational constraints \cite{astro1, astro2,astro3,measure1, measure2, measure3,measure4,measure5} the mass-radius relation displayed in Fig. \ref{Fig6} is very preliminary. In order to obtain more realistic predictions which could be contrasted with those presented  in Ref. \cite{fukushimatrace} one needs to further improve the NJL EoS by possibly incorporating strangeness as well as a hadronic piece to describe hybrid stars. Although this task is beyond the scope of the present work one may anticipate that using a running vector coupling, such as the one given here, will improve the NJL EoS.

\section{ Conclusions } 
This work shows that it is possible to describe  a non-conformal peak at $V_s^2 > 1/3$ using a standard effective quark model which contains a repulsive vector channel parametrized by a density dependent coupling. Here, it has been  suggested that the repulsion among (dressed) quarks  is important only up to the point where the chiral transition occurs  so that   repulsion among  (bare) quarks should be negligible.  To model this situation a 
 simple ansatz  was proposed. Basically, it interpolates between a regime where repulsion is high (the EoS is stiff) and a regime where repulsion low (the EoS is soft).   Thanks to this property  the model is able to predict a non-monotonic  behavior for $V_s^2$ which is in line with recent  estimates \cite{sinansimulation, michal, weisesimulation}.
 Regarding the conformal measure the results  indicate that  a non-conformal peak in $V_s^2$ is not in tension with the trace anomaly being positive for all densities, a result which agrees with a scenario proposed in  Ref. \cite{fukushimatrace}. This happens because the  model  generates a peak in $V_s^2$ at lower densities by stiffening  the EoS before the softening, at intermediate densities,    forces convergence towards the pQCD predictions. Moreover, a recent study shows  that    QCD predicts the softening of the EoS  in  most massive NSs \cite {new}. In a  crude application to the description of non-strange quark stars the model   predicts a smaller  maximum mass-radius ratio than the cases  where $G_V=0$ and $G_V=G_S/3$.  The results here obtained also allow us to conclude that although the standard NJL model, with a {\it fixed} $G_V$,  leads to stiffer EoS (an larger NS masses) \cite{jaziel} this is  accomplished by employing an EoS which  is in  disagreement with pQCD predictions at asymptotically high densities. 
 This seminal work can be further extended in order to describe more realistic situations. One possible such extension is to consider a diquark interaction channel in order to explore the high-density region of QCD. As far as the running of the vector coupling   is concerned one could also try different parameter values and or another {\it ansatz} to describe $G(\mu)$.

{\it Acknowledgments:} The author is partially supported by Conselho
Nacional de Desenvolvimento Cient\'{\i}fico e Tecnol\'{o}gico (CNPq),
Grant No  307261/2021-2  and by CAPES - Finance  Code  001.  
 This work has also been financed  in  part  by
Instituto  Nacional  de  Ci\^encia  e Tecnologia de F\'{\i}sica
Nuclear e Aplica\c c\~{o}es  (INCT-FNA), Process No.  464898/2014-5.

\end{document}